\documentclass[11pt]{article}

\usepackage[T1]{fontenc}
\usepackage[utf8]{inputenc}
\usepackage[margin=1in]{geometry}
\usepackage{lmodern}
\usepackage{microtype}
\usepackage{amsmath,amssymb,amsthm}
\usepackage{thm-restate}
\makeatletter
\ifcsname newcounteralias\endcsname
  \renewcommand\thmt@autorefsetup{%
    \@xa\def\csname\thmt@envname autorefname\@xa\endcsname\@xa{\thmt@thmname}%
  }%
\fi
\makeatother
\usepackage{mathtools}
\usepackage{array}
\usepackage{tabularx}
\usepackage[normalem]{ulem}
\usepackage[numbers,sort&compress]{natbib}
\usepackage[hidelinks]{hyperref}

\hypersetup{
  pdftitle={Strategic Tanking in Stable Matching Markets},
  pdfauthor={James P. Bailey}
}
\newtheorem{theorem}{Theorem}[section]
\newtheorem{lemma}[theorem]{Lemma}

\newtheorem{corollary}[theorem]{Corollary}
\theoremstyle{definition}
\newtheorem{definition}[theorem]{Definition}
\newtheorem{example}[theorem]{Example}
\theoremstyle{remark}

\newcommand{\SM}{\mathcal S}
\newcommand{\EQ}{\mathcal E}
\newcommand{\mM}{\mu_M}
\newcommand{\mW}{\mu_W}

\title{Strategic Tanking in Stable Matching Markets}
\author{James P. Bailey\\
Industrial and Systems Engineering\\
Rensselaer Polytechnic Institute\\
Troy, New York, USA\\
\texttt{bailej6@rpi.edu}}
\date{}

\begin{document}
\maketitle

\begin{abstract}
Motivated by a broad literature in psychology, organizational behavior, and matching markets, we introduce \textbf{strategic tanking}: an agent deliberately makes themselves less attractive to selected potential partners in order to improve their eventual assignment. Unlike traditional manipulation in stable matching, a tanker does not misreport whom they prefer; instead, they influence how others rank them. We show that tanking has a striking connection to classical preference manipulation. Cutoff tanking, in which an agent tanks with every potential partner below a chosen threshold, has exactly the same effect on the stable set as truncating one's own preference list at that threshold. Thus an agent can reproduce the power of a classical truncation while leaving their own preferences unchanged and every match acceptable. Arbitrary tanking, however, can go substantially further: under proposer-deferred acceptance, tanking equilibria can lie outside the sincere stable set. More broadly, whenever the sincere stable set contains multiple matchings, profitable strategic behavior remains available through either traditional preference manipulation or tanking, even when one side is restricted to honestly reporting their own preferences.
\end{abstract}


\medskip

\section{Introduction}
A favorable evaluation is usually an asset in an assignment process, but it can also expose an individual to an outcome they would rather avoid. Appearing less capable, less enthusiastic, or less compatible can then be purposeful behavior. Experiments on aversive assignments and organizational studies of negative impression management document such incentives \cite{KowalskiLeary1990,BeckerMartin1995}. Sandbagging similarly involves managing another person's
expectations downward~\cite{GibsonSachau2000,ShepperdSocherman1997}.

In a two-sided matching market, this behavior creates a strategic channel distinct from preference misreporting. Instead of falsely reporting whom they prefer, an agent can influence \emph{who prefers them}. We call this \emph{strategic tanking}: the agent moves themselves downward in selected opposite-side rankings while leaving their own ranking unchanged. 

This channel matters because standard strategyproofness addresses only how agents report their own preferences. With school priorities fixed, student-proposing DA protects students against profitable own-list manipulation \cite{AbdulkadirogluSonmez2003,DubinsFreedman1981}, but it does not prevent a school from influencing how students rank the school. More broadly, our results show that proposer-side strategyproofness does not make deferred acceptance free of strategic behavior. Whenever the sincere stable set contains multiple matchings, classical results imply that profitable traditional manipulation is available on the receiving side \cite{GaleSotomayor1985}, while we show that profitable tanking is available as well. Thus, even when one side is restricted to honestly reporting its own preferences, the outcome of DA can be strategically changed via strategic reporting or tanking.

\begin{example}[Students propose; a school tanks]\label{ex:admissions}
There are two students, $s_1,s_2$, and two schools, $A,B$, each with one seat. Rankings run from best to worst:
\[
\begin{array}{c@{\qquad}c}
\text{Students rank schools} & \text{Schools rank students}\\[2pt]
 s_1:A\succ B & A:s_2\succ s_1\\
 s_2:B\succ A & B:s_1\succ s_2.
\end{array}
\]
Student-proposing DA assigns $(s_1,A)$ and $(s_2,B)$. School $A$ can tank its interaction with student $s_1$, inducing $s_1:B\succ A$. Both students now propose to $B$, which retains $s_1$; $s_2$ then proposes to $A$. School $A$ obtains its preferred student without changing its own ranking. In contrast, permuting either school's complete own list cannot change the original outcome: the students initially propose to different schools and both proposals are accepted immediately.
\end{example}

Interviews provide a natural opportunity for this kind of influence. For example, the National Resident Matching Program (NRMP) uses an applicant-proposing algorithm, and its 2024 applicant survey identifies interview-day experience among frequently endorsed ranking considerations \cite{NRMPAlgorithm,NRMPApplicantSurvey2024}. Thus the same interaction used to learn a participant's preferences can also give the evaluated party an opportunity to shape them. Whether programs deliberately exploit this channel is a separate empirical question.

Although motivating applications such as residency matching and school
assignment are typically many-to-one, we develop the main theory in the
one-to-one setting. This restriction allows us to obtain the exact structural
and equilibrium characterizations that follow. However, the basic negative
phenomenon is not confined to one-to-one markets: profitable tanking and
tanking-induced sincere instability also arise in many-to-one settings.

\begin{table*}[!t]
\caption{Comparison of traditional preference manipulation and strategic tanking. Key differences are underlined.}
\label{tab:roadmap}
\small
\renewcommand{\arraystretch}{1.35}
\setlength{\tabcolsep}{5pt}
\begin{tabularx}{\textwidth}{>{\raggedright\arraybackslash}p{0.13\textwidth}>{\raggedright\arraybackslash}X>{\raggedright\arraybackslash}X}
\hline
\textbf{Result} & \textbf{Traditional Preference Manipulation} & \textbf{Strategic Tanking (Our Results)}\\
\hline
Profitable manipulation & If there are at least two stable matchings, some receiving-side agent can profitably manipulate \cite{GaleSotomayor1985}. & If there are at least two stable matchings, some receiving-side agent can profitably tank.\\

Optimal cutoff
&
Truncating at an agent's best stable partner attains that partner under any stable selection rule; same-side coalitions can attain these partners simultaneously.
&
Tanking at an agent's best stable partner attains that partner under any stable selection rule; same-side coalitions can attain these partners simultaneously.
\\

Cutoff filtering
&
Truncating below an attainable cutoff removes exactly the stable outcomes below that cutoff by shortening the agent's own list.
&
Tanking below the same attainable cutoff removes exactly the stable outcomes below that cutoff, \uline{while leaving the agent's own list unchanged}.
\\

Proposer protection & Under proposer-DA, proposing-side agents cannot benefit from own-list manipulation \cite{DubinsFreedman1981}. & Under proposer-DA, proposing-side agents cannot benefit from tanking.\\

Cutoff equilibria
&
Under cutoff truncation, a matching is an equilibrium outcome if and only if it is sincerely stable; under proposer-DA with non-strategic proposers, the unique equilibrium outcome is the sincere receiver-optimal stable matching.
&
Under cutoff tanking, a matching is an equilibrium outcome if and only if it is sincerely stable; under proposer-DA with non-tanking proposers, the unique equilibrium outcome is the sincere receiver-optimal stable matching.
\\

Unrestricted equilibria
&
Under proposer-DA with non-strategic proposers, receiving-side Nash outcomes are exactly the sincere stable matchings
\cite{GaleSotomayor1985}.
&
Under proposer-DA with non-tanking proposers, \uline{tanking equilibria can extend beyond the sincere stable set}.
\\

Same-side monotonicity & Relative to a fixed sincere baseline, beneficial receiving-side manipulation does not harm other receiving-side agents \cite{GonczarowskiFriedgut2013}. & \uline{Relative to every current state}, beneficial receiving-side tanking does not harm other receiving-side agents.\\

Profitable dynamics & Under standard own-list preference manipulation, strict best-response dynamics can cycle (Appendix~\ref{app:reportcycle}). & State-by-state monotonicity implies that every strict \uline{better-response tanking path reaches equilibrium}.\\
\hline
\end{tabularx}
\end{table*}

\subsection{Contributions and roadmap}
The paper studies strategic tanking---intentionally lowering one's position in others' rankings---and follows the classical progression from Gale and Shapley \cite{GaleShapley1962}, through Dubins and Freedman \cite{DubinsFreedman1981}, to Gale and Sotomayor \cite{GaleSotomayor1985}. We ask the corresponding questions for this new strategic action: what can an individual gain by tanking, what can coordinated tanking behavior accomplish, and which outcomes can arise in equilibrium when individuals tank? Table~\ref{tab:roadmap} summarizes the main parallels to traditional strategic behavior.

Table~\ref{tab:roadmap} also reveals a broader strategic message. The familiar proposer--receiver asymmetry of deferred acceptance survives when strategic behavior is expanded beyond preference reports. Proposers cannot benefit from traditional own-list manipulation or tanking. Yet, profitable receiving-side behavior exists under both strategic channels: classical preference manipulation and strategic tanking. Restricting one side to honest preference reports does not eliminate strategic behavior.

The central structural contribution is a simple filter. Fix an agent and a target partner $t$ that the agent can receive in some stable matching. If the agent tanks with every potential partner ranked below $t$---a strategy we call $t$-cutoff tanking---the surviving stable matchings are exactly those assigning the agent a partner at least as good as $t$. This is precisely the stable-set effect of traditional truncation below $t$, despite the tanker leaving her own list unchanged and keeping every pair acceptable. Choosing $t$ to be the agent's best current stable partner is optimal under any stable selection rule, and the same idea extends to same-side coalitions. Section~\ref{sec:power} develops these local, largely mechanism-independent results.

Section~\ref{sec:equilibrium} turns from what tanking can accomplish to the outcomes that arise when agents use it strategically. Cutoff tanking remains tightly connected to sincere stability: bilateral cutoff equilibria support exactly the sincere stable matchings, while one-sided cutoff equilibrium selects the strategic side's sincere-optimal stable matching. Arbitrary tanking is less benign. Even under proposer-DA with non-tanking proposers, tanking equilibria can be unstable under the original preferences. We nevertheless obtain a strong same-side monotonicity property, finite convergence of strict better responses, and a complete characterization of the broader equilibrium outcome set. The contrast identifies the equilibrium-selection questions raised by strategic preference formation.

\section{Behavioral Motivation and Related Work}\label{sec:related}
\subsection{Why an agent might want to look worse}
\emph{Negative impression management} is the closest behavioral connection. In job-simulation experiments, Kowalski and Leary find unfavorable self-presentation when appearing better adjusted exposes participants to an onerous assignment \cite{KowalskiLeary1990}. Becker and Martin study reported employment experiences from 162 respondents \cite{BeckerMartin1995}. Their categories include reduced performance, working below one's capabilities, withdrawal, and unfavorable attitudes; avoidance motives include unwanted transfers or promotions and work with particular people or groups.

Related behavior has been studied through feigned incompetence and lowered expectations \cite{ShepperdSocherman1997,GibsonSachau2000}, organizational knowledge hiding through \emph{playing dumb} \cite{ConnellyEtAl2012}, sporting contexts involving handicap manipulation and impressions \cite{SachauEtAl2014}, and laboratory labor markets in which current performance affects future targets \cite{CharnessKuhnVilleval2011}. McLuhan describes a ``cloak of incompetence'' \cite{McLuhan2020}; Stadnicka develops a research agenda around feigned digital incompetence \cite{Stadnicka2024}; and reviews place sandbagging and intentional task underperformance in a broader research landscape \cite{ProdanBoza2025,WinchesterEtAl2026}. The feature relevant here is deliberate deterioration of one's evaluation by a potential partner when a different acceptable assignment is preferred.

\subsection{Opportunities for tanking in stable matching markets}
A growing stable matching literature relaxes the classical assumption that complete preference lists are known before the mechanism is run. Preferences may instead be learned through interviews, elicitation, noisy observations, or repeated interaction \cite{RastegariEtAl2013,RastegariGoldbergManlove2016,AshlagiEtAl2025,AthanasopoulosGeorgeDimitrakakis2025,HosseiniRoyZhang2024,AthanasopoulosDimitrakakis2026}. These models make explicit that the rankings supplied to a stable matching mechanism can be the output of an information-acquisition process rather than an exogenous primitive.

Those same learning processes create the opportunity studied here. When an agent's ranking is inferred from an observed interaction, the agent being evaluated may be able to influence the resulting signal. An agent who prefers to avoid a particular match may deliberately perform poorly, appear disinterested, or otherwise make themselves less attractive. Residency interviews make the opportunity concrete: applicants use interview experiences when forming program rankings \cite{NRMPApplicantSurvey2024}. The evidence establishes that interviews help form rankings, not that programs deliberately tank.

The stable-matching literature also gives a reason such behavior can pay. Balinski and S\"onmez formalize mechanisms that \emph{respect improvements}: improving an agent's position in opposite-side rankings should not worsen that agent's assignment \cite{BalinskiSonmez1999}. Hatfield, Kojima, and Narita study related attractiveness changes with fixed school preferences and discuss institutions benefiting when students become less interested in them \cite{HatfieldKojimaNarita2016}. 
Klijn, Pais, and Vorsatz experimentally study this comparative
static with passive schools, including instances in which a school
rises in students' preferences while its student-optimal
stable match worsens~\cite{KlijnPaisVorsatz2020}.
Reversing such an improvement gives precisely the comparative-static prerequisite for profitable tanking. These papers do not model deliberate tanking as a strategic game; rather, they show that stable matching mechanisms can create incentives to become less preferred.

\subsection{Strategic behavior in stable matching}
Classical reporting analysis establishes the proposer/receiver incentive asymmetry and receiving-side manipulation through falsification and truncation \cite{DubinsFreedman1981,Roth1982,GaleSotomayor1985}. Complete-list permutations and minimal-blacklist constructions further distinguish reordering from deleting acceptable partners \cite{TeoSethuramanTan2001,ShenDengTang2021,Gonczarowski2014Blacklists}.

Rather than allowing agents to alter their own submitted preferences, we study a different strategic channel: agents keep their own preference lists fixed and instead strategically lower their position in selected agents' rankings. Thus, strategic tanking shifts manipulation from how an agent reports their own preferences to how that agent influences the preferences held by the opposite side.

The closest prior work to our strategic model is accomplice manipulation. Hosseini, Umar, and Vaish study settings in which an agent benefits when another participant strategically alters their reported preference list on the beneficiary's behalf \cite{HosseiniUmarVaish2021,HosseiniUmarVaish2022}; more recent work examines the resulting strategic games \cite{HosseiniLisowskiPathak2025}. Like tanking, accomplice manipulation can improve an agent's assignment by changing rankings on the opposite side of the market. The strategic control is nevertheless different: accomplice manipulation requires cooperation from another agent who misreports their own preferences, whereas tanking gives the beneficiary direct control over selected downward changes in how they are ranked while leaving their own preference list fixed.


\section{Preliminaries and Strategic Tanking}\label{sec:preliminaries}
\subsection{Sincere and induced preferences}
Let $M$ and $W$ be disjoint sets of $n$ agents, called men and women following the classical notation. Every agent strictly ranks all members of the opposite side, and every pair is acceptable. The \emph{sincere} profile $P=(P_M,P_W)$ records the underlying full-information rankings that would result from sincere evaluation. Like the classical setting, strategic behavior may instead produce a current or \emph{induced} profile $Q=(Q_M,Q_W)$.

A matching $\mu$ is a bijection between $M$ and $W$; we use $\mu$ in both directions. Write $a\succ_i^Q b$ for agent $i$'s comparison at $Q$, omitting the superscript when $Q=P$. A pair $(m,w)$ blocks $\mu$ at $Q$ if both prefer each other to their assigned partners. Let $\SM(Q)$ denote the stable set. Membership in $\SM(P)$ is \emph{sincere stability}; membership in $\SM(Q)$ is \emph{induced stability}.

Stable-matching results guarantee two extremal matchings when there are at least two stable matchings. There is a stable matching $\mM(Q)$ that every man weakly prefers to every other stable matching and that is simultaneously weakly worst for every woman; symmetrically, $\mW(Q)$ is simultaneously best for every woman and worst for every man \cite{GaleShapley1962,GaleSotomayor1985}. We refer to these as the men-optimal and women-optimal stable matchings. When one side is designated the proposing or receiving side, we use the analogous proposer-optimal and receiver-optimal terminology.

Our analysis extends to unbalanced one-to-one markets and incomplete preference lists through the standard introduction of dummy agents. General many-to-one matching requires additional analysis.  However, the negative
possibility results below extend immediately by embedding the one-to-one
examples.

A \emph{stable selection rule} is any rule that returns some matching in $\SM(Q)$ at every profile $Q$. Deferred acceptance (DA) is the canonical example. Under proposer-DA, unmatched proposers apply down their current rankings and receivers tentatively hold their favorite offer; the resulting matching is the proposer-optimal stable matching. Reversing the proposing side produces the opposite extreme \cite{GaleShapley1962}. We use proposer-DA when the particular selection rule matters and otherwise work with arbitrary stable selection rules.

\subsection{Arbitrary and Cutoff Tanking}
A key difference between our model and the classical strategic analysis is the action available to agents. Traditional manipulation allows an agent to change the preference list they submit. We instead consider agents who keep their own preferences fixed but strategically reduce how highly they are ranked by agents on the opposite side. We refer to such downward self-demotions as \emph{tanking}.

We consider two classes of tanking actions. The first is arbitrary tanking, in which an agent may selectively lower their position in any collection of opposite-side rankings. This is the broad strategy space used in our unrestricted equilibrium analysis. The second is \(t\)-cutoff tanking, a structured subclass in which the agent tanks with every partner ranked below a target \(t\). Cutoff tanking plays a central role because it admits an exact stable-set characterization and closely parallels truncation in the classical manipulation model.

\begin{definition}[Unilateral tanking]\label{def:tanking}
Fix a current profile $Q$ and an agent $i$. A profile $Q'$ is obtainable by \emph{tanking by $i$ at $Q$} if, in any selected opposite-side lists, $i$ is moved to the same or a lower position while the relative order of every other agent in those lists is preserved. Agent $i$'s own list does not change. Let $\mathcal T_i(Q)$ denote the profiles obtainable by such an action.
\end{definition}

\begin{definition}[$t$-cutoff tanking]\label{def:tcutoff}
Fix a current profile $Q$, an agent $i$, and a partner $t$ assigned to $i$ in some matching in $\SM(Q)$. A \emph{$t$-cutoff tanking action} moves $i$ to the bottom of every opposite-side agent's list corresponding to a partner that $i$ ranks below $t$ in $Q_i$. We denote the resulting profile by $Q^{i,t}$.
\end{definition}

An admissible tanking history is a finite sequence of current-state tanking actions.
We use a complete-information benchmark for strategic calculation: a strategic agent knows the relevant current profile and the stable selection rule. When later discussing equilibrium, we distinguish whether welfare is evaluated according to the original sincere ranking or the current induced ranking. Under the one-sided DA model that forms our main arbitrary-equilibrium analysis, the strategic side's own rankings never change and the distinction disappears.

\section{The Power of Strategic Tanking}\label{sec:power}
The core results in this section concern how tanking changes the stable set. They are therefore independent of the matching algorithm or stable selection rule, except where a DA-specific consequence is stated explicitly. All statements are local: the current profile $Q$ may itself be the product of earlier strategic actions.

\subsection{Individual tanking}\label{sec:individual}
We begin by establishing how an individual tanking action can change the set of stable outcomes. The adjacent-swap property of Bailey and Tovey \cite[Property~3]{BaileyTovey2025} identifies which assignments can gain or lose stability.

\begin{restatable}[{Bailey and Tovey 2025~\cite{BaileyTovey2025}}]{lemma}{localstablelemma}\label{lem:local}
Suppose agent $i$ moves downward in an opposite-side list, crossing a set $X$. If a matching loses stability, that list owner was matched to $i$; if a matching gains stability, that list owner is matched to an agent in $X$.
\end{restatable}

Lemma~4.1 shows that tanking can change the stable set in both directions: it may remove previously stable matchings, but it may also allow previously unstable matchings to become stable. The following example shows that this latter possibility can actually occur. We return to the same example in Section~\ref{sec:equilibrium} when studying arbitrary tanking equilibria.

\begin{example}[A profitable move with a sincere blocking pair]\label{ex:unstable}
Consider these complete rankings, from best to worst:
\[
\begin{array}{c@{\quad}c}
 m_1:w_3,w_2,w_1 & w_1:m_1,m_2,m_3\\
 m_2:w_2,w_1,w_3 & w_2:m_3,m_1,m_2\\
 m_3:w_1,w_2,w_3 & w_3:m_3,m_2,m_1.
\end{array}
\]
Men-proposing DA produces $(m_1,w_3),(m_2,w_2),(m_3,w_1)$. Woman $w_2$ can profitably tank by moving below $w_1$ at $m_2$ and below $w_3$ at $m_3$. The new DA pairs are $(m_1,w_2),(m_2,w_1),(m_3,w_3)$, so $w_2$ improves from $m_2$ to $m_1$.

The new matching is stable under the induced post-tanking preferences, as guaranteed by DA, but it is not stable under the original sincere preferences: $(m_3,w_2)$ is a sincere blocking pair. Thus a profitable tanking move can create a sincerely unstable outcome. We later establish this outcome is a tanking equilibrium.
\end{example}

\paragraph{Many-to-one extension.}
The negative phenomenon in Example~\ref{ex:unstable} is not specific to
one-to-one matching.  Unit-capacity institutions give the result immediately
as a special case of many-to-one matching.  Moreover, for any larger positive
capacities, the same example can be embedded by assigning each institution
$q-1$ dedicated agents whom it ranks above the agents in the example and who
rank that institution first.  These agents occupy the additional seats
throughout, while the remaining seat reproduces the one-to-one instance.
Hence a profitable tanking action can also produce an outcome that is stable
under the induced preferences but blocked under the sincere preferences in a
many-to-one market.

Cutoff tanking behaves differently. If the tanker chooses a target stable partner $t$ and tanks with everyone they rank below $t$, the action filters the current stable set and creates no new matchings.

\begin{restatable}[Exact cutoff filtering]{theorem}{exactcutofffilter}\label{thm:filter}
Fix a current profile $Q$, an agent $i$, and a stable partner $t$ of $i$ at $Q$. Then
\[
\SM(Q^{i,t})=\{\mu\in\SM(Q):\mu(i)\succeq_i^Q t\}.
\]
\end{restatable}

The proof is in Appendix~\ref{app:filter}. Intuitively, matchings meeting the cutoff survive by Lemma~\ref{lem:local}; a stable matching assigning $i$ to $t$ prevents the opposite-side-optimal extreme from crossing below the cutoff, so no new stable matching can appear.

Interestingly, this filter is exactly the one produced by truncation in the classical manipulation model. Let $Q^{i,\le t}$ be the profile in which only $i$ changes its own list, retaining exactly the partners ranked at least as high as $t$ in the current order.

\begin{restatable}[Exact equivalence with traditional truncation]{corollary}{truncationequivalence}\label{cor:truncation}
For every current stable partner $t$ of $i$,
\[
\SM(Q^{i,t})=\SM(Q^{i,\le t})
=\{\mu\in\SM(Q):\mu(i)\succeq_i^Q t\}.
\]
Thus cutoff tanking and cutoff truncation apply the same filter.
\end{restatable}

The proof is in Appendix~\ref{app:filter}.
This is the first major parallel in Table~\ref{tab:roadmap}. Gale and Sotomayor use truncation at a best stable partner \cite{GaleSotomayor1985}; cutoff tanking reproduces the same stable-set effect while the actor's own list and acceptability remain unchanged. Because Theorem~\ref{thm:filter} is local, the equivalence can be applied repeatedly along a cutoff history.

For later use, let
\[
b_i(Q)=
\begin{cases}
\mM(Q)(i),&i\in M,\\
\mW(Q)(i),&i\in W.
\end{cases}
\]
denote agent $i$'s best partner among the stable matchings of the current profile $Q$. Thus $b_i(Q)$ is the partner assigned to $i$ by the stable matching optimal for $i$'s side at the current profile.

Like its truncation analogue, cutoff tanking at $b_i(Q)$ is optimal; no arbitrary unilateral tanking action can improve on this outcome.

\begin{restatable}[No tanking beyond the best stable partner]{theorem}{unilateralupper}\label{thm:upper}
Fix a current profile $Q$ and agent $i$. If $Q'\in\mathcal T_i(Q)$ and $\eta\in\SM(Q')$, then
\[
\eta(i)\preceq_i^Q b_i(Q).
\]
\end{restatable}

The full proof is in Appendix~\ref{app:upper}. 
We show that if tanking allowed $i$ to obtain a partner preferred to $b_i(Q)$, we could
construct an auxiliary instance in which the tanking is undone and $i$
instead reports that partner first on their own list.  The post-tanking
matching remains stable in this auxiliary instance, so proposer-optimal DA
would assign $i$ that partner.  This would constitute a profitable
own-list manipulation by a proposer, contradicting the strategyproofness
result of Dubins--Freedman \cite{DubinsFreedman1981}.

Thus cutoff tanking attains the stable-partner upper bound, paralleling the
classical search for optimal preference manipulations
\cite{TeoSethuramanTan2001}, while leaving the actor's own preference list
unchanged.

\begin{corollary}[Cutoff tanking is an optimal best response]\label{cor:bestresponse}
Consider any rule that selects a stable matching of the current profile. For any agent $i$, cutoff tanking at $b_i(Q)$ is a best response among all unilateral tanking actions at $Q$; every stable selection after that cutoff assigns $i$ the partner $b_i(Q)$.
\end{corollary}

The same best-response conclusion holds for traditional cutoff truncation because Corollary~\ref{cor:truncation} gives the identical stable-set filter. This provides a direct strategic, rather than merely structural, parallel between the two cutoff actions.

\begin{corollary}[DA proposers are tanking-proof]\label{cor:proposerproof}
Under proposer-DA, no proposing-side agent can improve by unilateral tanking at the current profile.
\end{corollary}

A proposer already receives its best current stable partner. This exactly mirrors the classical proposer-side strategyproofness result for own-list manipulation \cite{DubinsFreedman1981}. It will justify treating DA proposers as non-strategic when we analyze arbitrary tanking in Section~\ref{sec:arbitraryeq}. The optimal cutoff result also suggests a natural equilibrium filtering process; we turn to equilibrium in Section~\ref{sec:equilibrium}.

\subsection{Coordinated tanking}\label{sec:coordinated}
We next consider how coordinated tanking changes the stable set and extend the individual results of Section~\ref{sec:individual} to same-side coalitions. Throughout this subsection, the coalition lies entirely in $M$ or entirely in $W$. This restriction matters: with a mixed coalition, one member's tanking can alter another member's own induced ranking and therefore change that member's cutoff target. We defer such bilateral feedback to Section~\ref{sec:equilibrium} and future work. 
All proofs in this section follow identically to their single-agent tanking counterparts and are deferred to the appendix. 

Fix a current profile $Q$, a same-side coalition $C$, and stable-partner cutoffs $t_i$ that are \emph{jointly feasible}: some $\nu\in\SM(Q)$ satisfies $\nu(i)\succeq_i^Q t_i$ for every $i\in C$. The individual filtering theorem extends by intersecting the coalition members' cutoff requirements.

\begin{restatable}[Joint cutoff filtering]{theorem}{jointcutofffilter}\label{thm:coalitionfilter}
After the members of $C$ apply their cutoffs, in any order,
\[
\SM(Q^{C,t})
=\{\mu\in\SM(Q):\mu(i)\succeq_i^Q t_i\ \text{for all }i\in C\}.
\]
The result is independent of the order in which the coalition's cutoff actions are applied.
\end{restatable}

 The statement is the direct same-side analogue of Theorem~\ref{thm:filter}, and coordinated truncation at the same jointly feasible cutoffs produces the same stable set.

The unilateral upper bound also has a coordinated counterpart. It is necessarily an all-members statement: coordinated tanking can sometimes help one coalition member beyond the original stable frontier if another member does not strictly benefit.

\begin{restatable}[Coalitional upper bound]{theorem}{coalitionalupper}\label{thm:coalitionupper}
Fix a current profile $Q$ and a same-side coalition $C$. No coordinated tanking action by $C$ can produce an induced stable matching $\eta$ satisfying
\[
\eta(i)\succ_i^Q b_i(Q)\qquad\text{for every }i\in C.
\]
\end{restatable}

Together with joint filtering, this gives the coordinated analogue of the optimal cutoff response.

\begin{corollary}[Optimal coordinated cutoff tanking]\label{cor:coalitionbest}
For any current profile $Q$ and same-side coalition $C$, the cutoffs $t_i=b_i(Q)$ are jointly feasible. Applying them simultaneously guarantees every coalition member its best current stable partner under any stable selection rule. No coordinated tanking action can make every coalition member strictly better than this cutoff outcome.
\end{corollary}

The side-optimal stable matching simultaneously assigns every same-side coalition member $b_i(Q)$, so the targets are jointly feasible. This mirrors coordinated truncation: within the cutoff strategy class the same stable-set operator yields the same simultaneous optimum. Equilibrium consequences are postponed to the next section.

\section{Equilibrium}\label{sec:equilibrium}

Section~\ref{sec:power} established what strategic tanking can accomplish: tanking can reshape the stable set, and cutoff tanking provides optimal individual and coordinated responses under broad classes of stable selection rules. We now turn from the \emph{power} of tanking to the outcomes that arise when agents use these actions strategically. We first study equilibria under cutoff tanking, where the filtering structure yields strong stability guarantees, and then consider arbitrary tanking, where substantially different outcomes can arise.

Before examining these outcomes, we address a conceptual issue that arises when both sides tank. Tanking changes how an agent is ranked by the opposite side, so under bilateral tanking an agent's own current ranking may itself have been altered by earlier strategic behavior. We therefore introduce two interpretations, which we call the \emph{fixed-preference} and \emph{endogenous-preference} interpretations.

Under the \emph{fixed-preference interpretation}, agents evaluate welfare and strategic choices using their original sincere rankings $P_i$, even when their induced rankings have changed. Under the \emph{endogenous-preference interpretation}, agents instead evaluate subsequent choices using their current rankings $Q_i$. These interpretations concern how agents evaluate strategic deviations; the matching mechanism itself continues to act on the current profile $Q$.

Under the endogenous-preference interpretation, a stable selection rule always returns a matching that is stable with respect to the final induced profile $Q$. The more informative question is therefore how strategic behavior changes the outcome relative to the original sincere profile $P$. In particular, a matching that is stable under strategically induced preferences may fail to be stable under the original preferences. This distinction may be especially important when induced evaluations are temporary or when agents gradually recover information about their underlying untanked preferences.

For cutoff tanking, the distinction between the two interpretations is largely benign: cutoff actions only remove stable matchings, and the surviving stable alternatives preserve the relevant preference comparisons. When we later turn to arbitrary tanking, we specialize to proposer-deferred acceptance with non-tanking proposers. In that setting only the receiving side acts strategically, so the strategic agents' own preferences remain unchanged and the fixed- and endogenous-preference interpretations coincide.

\subsection{Equilibria under Cutoff Tanking}\label{sec:cutoffeq}
We begin with cutoff tanking, where the equilibrium picture is especially clean. Because every cutoff action only filters the existing stable set, cutoff strategies preserve sincere stability. For any stable selection rule, the outcomes supportable at bilateral cutoff equilibria are exactly the matchings stable under the original profile $P$. If only one side is strategic, cutoff equilibrium selects that side's sincere-optimal stable matching.

\begin{definition}[Cutoff equilibrium]\label{def:cutoffeq}
Fix a stable selection rule, a set of strategic agents, and a profile $Q$ reachable from sincere $P$ through current-state cutoff tanking at current stable partners. The profile $Q$ is a \emph{fixed-preference cutoff equilibrium} if no strategic agent can obtain an assignment it strictly prefers according to $P_i$ through an additional current-state cutoff action. It is an \emph{endogenous-preference cutoff equilibrium} if the same condition holds with outcomes evaluated according to $Q_i$.
\end{definition}

The following technical observation explains why the two interpretations agree on cutoff-reachable stable alternatives.

\begin{restatable}[Surviving-partner order invariance]{lemma}{survivingorder}\label{lem:survivingorder}
Let $Q$ be reachable from sincere $P$ through a sequence of current-state cutoff tanking actions at current stable partners. For every agent $i$ and every $\mu,\nu\in\SM(Q)$,
\[
\mu(i)\succ_i^P\nu(i)
\quad\Longleftrightarrow\quad
\mu(i)\succ_i^Q\nu(i).
\]
\end{restatable}

The proof is in Appendix~\ref{app:cutoffeq}. Intuitively, whenever a cutoff changes a comparison in another agent's list, the tanker simultaneously eliminates every surviving stable matching in which that altered comparison could determine the other agent's partner.

\begin{restatable}[Bilateral cutoff equilibrium]{corollary}{bilateralcutoffeq}\label{cor:cutoffsingleton}
Fix any stable selection rule. Let $Q$ be cutoff-reachable from sincere $P$, allow both sides to make additional cutoff deviations, and evaluate welfare under either the fixed- or endogenous-preference interpretation. Then $Q$ is a cutoff equilibrium if and only if  $\SM(Q)$ is a singleton.
\end{restatable}

This is an immediate lattice consequence. If the selected matching differs from $\mM(Q)$, some man prefers his men-optimal stable partner and can force it by cutoff; symmetrically, if it differs from $\mW(Q)$, some woman can profitably cutoff. Lemma~\ref{lem:survivingorder} makes the same comparison valid under fixed preferences. A singleton stable set leaves no cutoff improvement.

The outcome-level statement is more useful: cutoff equilibria recover exactly the sincere stable set and therefore always exist.

\begin{restatable}[Cutoff equilibria recover the sincere stable set]{theorem}{cutoffsupport}\label{thm:cutoffsupport}
Fix any stable selection rule. Starting from sincere $P$, allow both sides to use only current-state cutoff tanking at current stable partners. Under either the fixed- or endogenous-preference interpretation, a matching $\mu$ is the outcome of a cutoff equilibrium if and only if $\mu\in\SM(P)$. In particular, a cutoff equilibrium always exists.
\end{restatable}

The proof is in Appendix~\ref{app:cutoffeq}. Every cutoff action contracts the current stable set, so no cutoff history can create a sincerely unstable outcome. Conversely, every sincere stable matching can be isolated by a finite sequence of cutoffs and hence supported at equilibrium. Because tanking and truncation apply the same cutoff filter, this equilibrium-level correspondence has a direct traditional-truncation analogue. It also mirrors Gale and Sotomayor's conclusion that suitable receiving-side reports can support every sincere stable matching \cite{GaleSotomayor1985}, though the strategy spaces differ.

With only one strategic side, the prediction becomes substantially sharper.
Moreover, the fixed- and endogenous-preference interpretations coincide in this
setting: tanking by the strategic side changes only the opposite side's
rankings, so the strategic agents' own preferences remain equal to their
sincere preferences throughout the process.

\begin{restatable}[One-sided cutoff equilibrium selects the side-optimal matching]
{theorem}{onesidecutoff}\label{thm:onesidecutoff}
Fix any stable selection rule and sincere profile $P$. Suppose one side does not
tank, while agents on the opposite side may use current-state cutoff tanking.
Then a cutoff equilibrium exists, and every cutoff equilibrium yields the
sincere stable matching optimal for the strategic side.
\end{restatable}

The proof is in Appendix~\ref{app:cutoffeq}. 
Theorem~\ref{thm:onesidecutoff} is independent of the stable matching mechanism. Under proposer-deferred acceptance, this
assumption is particularly natural because proposers cannot benefit from
tanking.

\begin{corollary}[Cutoff equilibrium under proposer-DA]
\label{cor:da-cutoff-equilibrium}
Under proposer-deferred acceptance, proposers cannot benefit from tanking.
Thus, when proposers do not tank, every receiving-side cutoff equilibrium
yields the sincere receiver-optimal stable matching.
\end{corollary}

This mirrors the classical strategic analysis of deferred acceptance: the
proposing side has no incentive to manipulate, and once proposers remain
non-strategic, strategic behavior by the receiving side selects its
side-optimal stable outcome.


Taken together, the results of this subsection show that cutoff tanking is
particularly well behaved.  Bilateral cutoff equilibria recover exactly the
sincere stable set, while one-sided cutoff equilibrium selects the strategic
side's sincere-optimal stable matching.  We next ask what remains true when the
receiving side is no longer restricted to cutoff actions.  As the next
subsection shows, allowing arbitrary tanking substantially enlarges the set of
possible equilibrium outcomes and can even destroy stability with respect to
the sincere preferences.

\subsection{Equilibria under Arbitrary Tanking}\label{sec:arbitraryeq}

We now remove the cutoff restriction and allow the receiving side to tank
arbitrarily.  The equilibrium picture becomes substantially less selective.
Throughout this subsection, we specialize to proposer-deferred acceptance with
non-tanking proposers.  In our notation, men propose and do not tank, while
women may make arbitrary self-demotions, so every current profile has the form
$Q=(Q_M,P_W)$.  Because the strategic side's own preferences remain unchanged,
the fixed- and endogenous-preference interpretations coincide.

\begin{definition}[Tanking equilibrium]\label{def:equilibrium}
Under men-proposing DA with non-tanking men, a profile $Q=(Q_M,P_W)$ is a
\emph{tanking equilibrium} if no woman can obtain a strictly better sincere
partner through any further tanking action:
\[
\mM(Q)(w)\succeq_w \mM(Q')(w)
\qquad
\text{for all }w\in W,\ Q'\in\mathcal T_w(Q).
\]
\end{definition}

We first characterize the equilibria admitted by this broader strategy space.
The following result reduces the equilibrium condition to a familiar property
of the induced stable-matching problem.

\begin{restatable}[DA tanking equilibrium and unique induced stability]
{theorem}{uniqueinduced}\label{thm:equilibriumunique}
Under men-proposing DA with non-tanking men, a profile $Q=(Q_M,P_W)$ is a
tanking equilibrium if and only if $\SM(Q)$ is a singleton.
\end{restatable}

If $\SM(Q)$ is non-singleton, some woman prefers her induced woman-optimal
partner to the men-proposing outcome and can force that partner through cutoff
tanking.  If $\SM(Q)$ is a singleton, Theorem~\ref{thm:upper} rules out any
better induced-stable outcome following further tanking.  In particular,
tanking equilibria always exist: the receiver-optimal cutoff construction of
Section~\ref{sec:cutoffeq} produces a profile with a unique induced stable
matching and is therefore also an equilibrium against arbitrary tanking.

\paragraph{Which outcomes can tanking equilibria support?}
Theorem~\ref{thm:equilibriumunique} turns the unrestricted equilibrium problem
into an implementation question: which assignments can the receiving side make
uniquely stable by lowering itself in the proposers' rankings?  The next
theorem characterizes this outcome set exactly.

\begin{restatable}[Characterization of supportable tanking-equilibrium outcomes]
{theorem}{supporttheorem}\label{thm:support}
Let $\EQ(P)$ be the set of outcomes supportable at a tanking equilibrium
obtainable from $P$ by some admissible sequence of arbitrary tanking actions
under men-proposing DA with non-tanking men. Then
\[
\begin{aligned}
\EQ(P)=\{\mW(R_M,P_W):\;&R_M\text{ is any strict complete}\\
&\text{profile of men's rankings}\}.
\end{aligned}
\]
Membership of a target matching in this set, and a supporting
tanking-equilibrium profile when one exists, can be determined in $O(n^2)$
time.
\end{restatable}

The proof of Theorem~\ref{thm:support} is given in Appendix~\ref{app:support}.
Theorem~\ref{thm:support} has two contrasting implications.  On the one hand,
arbitrary tanking gives the receiving side considerable strategic control:
every supportable equilibrium outcome is the receiving-side-optimal stable
matching for \emph{some} induced profile of proposing-side preferences.  Thus
the strategic side does more than influence which stable matching is selected
from a fixed market; through tanking, it can shape which proposing-side
preference profile the mechanism effectively faces.  On the other hand, that
induced profile need not preserve the stable outcomes of the sincere market, so
this additional control can come at the cost of sincere stability.

Indeed, after exchanging the roles of the two sides, the right-hand side of
Theorem~\ref{thm:support} is precisely the complete-list implementation set
studied by Kobayashi and Matsui
\cite{KobayashiMatsui2009,KobayashiMatsui2010}.
Their results therefore provide an $O(n^2)$ membership test and construction
of a supporting proposing-side profile; Appendix~\ref{app:support} uses this
criterion for the algorithmic part of the theorem.

The characterization also lets us assess how restrictive tanking equilibrium
actually is.  We next examine how large the supportable outcome set can be and
how it relates to the receiving side's Pareto frontier.  The first corollary
shows that the equilibrium set can range from a singleton to the entire set of
perfect matchings; the second shows that it always contains every
receiving-side Pareto-efficient perfect matching.
Corollary~\ref{cor:extremes} follows directly from
Theorem~\ref{thm:support} and Corollary~\ref{cor:pareto} is proved in
Appendix~\ref{app:pareto}.

\begin{corollary}[Extremes of arbitrary-tanking equilibrium support]
\label{cor:extremes}
The tanking-equilibrium outcome set can range from a singleton to all $n!$
perfect matchings.
\begin{enumerate}
    \item If all women have the same sincere ranking of men, every perfect
    matching lies in $\EQ(P)$.
    \item If all women have distinct sincere first choices, $\EQ(P)$ contains
    only the matching assigning every woman her first choice, namely $\mW(P)$.
\end{enumerate}
\end{corollary}

The support characterization above describes which assignments can arise at a
tanking equilibrium, but it does not by itself indicate how this set relates to
the receiving side's welfare.  A natural benchmark is Pareto efficiency.  We
say that a perfect matching $\mu$ is \emph{receiving-side Pareto efficient} if
there is no perfect matching $\nu$ such that every receiving-side agent weakly
prefers $\nu$ to $\mu$ and at least one receiving-side agent strictly prefers
$\nu$.

The next result shows that the tanking-equilibrium outcome set is broad enough
to contain the entire receiving-side Pareto frontier.  Thus every assignment
that cannot be jointly improved for the receiving side can be supported at a
tanking equilibrium, even though such an assignment need not be stable under
the sincere preferences.

\begin{restatable}[Pareto-efficient outcomes and beyond]
{corollary}{paretocor}\label{cor:pareto}
Every perfect matching that is Pareto efficient for the women lies in
$\EQ(P)$, but the reverse inclusion fails: some tanking-equilibrium outcomes
are woman-Pareto inefficient.
\end{restatable}

The inclusion is the substantive part of Corollary~\ref{cor:pareto}: arbitrary
tanking can support every assignment on the receiving side's Pareto frontier.
The reverse inclusion is less surprising.  By the classical lattice structure,
the sincere receiver-optimal stable matching weakly Pareto dominates every
other sincere stable matching from the receiving side's perspective
\cite{GaleShapley1962,GaleSotomayor1985}.  Hence whenever the sincere stable set
contains more than one matching, some sincere stable matchings are already not
receiving-side Pareto efficient.  Since sincere stable matchings are themselves
supportable through cutoff tanking, the tanking-equilibrium outcome set can
contain Pareto-inefficient assignments even without leaving the sincere stable
set.

Together, Theorem~\ref{thm:support} and
Corollaries~\ref{cor:extremes}--\ref{cor:pareto} show that arbitrary tanking
admits a broad equilibrium set.  It contains every receiving-side
Pareto-efficient perfect matching, can range from a singleton to all $n!$
perfect matchings, and may also contain Pareto-inefficient outcomes.  More
importantly for stability, unlike the cutoff equilibria of
Section~\ref{sec:cutoffeq}, arbitrary-tanking equilibria need not remain in the
sincere stable set.

This broad supportability result raises a natural equilibrium-selection
question.  Perhaps the less desirable equilibria require implausible strategic
histories---for example, agents deliberately taking actions that make
themselves worse off.  We therefore consider a simple behavioral refinement:
starting from the sincere profile, require every tanking action along the
history to be strictly profitable for its mover.

\paragraph{Profitable tanking as an equilibrium refinement.}
Unfortunately, restricting attention to profitable tanking does not restore stability.

\begin{restatable}[Profitable tanking can yield sincere instability]
{proposition}{unstableeq}\label{prop:unstableeq}
Under men-proposing DA with non-tanking men, there exists a sincere profile
$P$ and a sequence of strictly profitable receiving-side tanking actions that
reaches a tanking equilibrium whose outcome is not stable with respect to
$P$.  In fact, such an equilibrium can be reached from $P$ after a single
strictly profitable tanking action.
\end{restatable}

Example~\ref{ex:unstable} provides the construction, and
Appendix~\ref{app:unstable} verifies both the equilibrium and the sincere
blocking pair.  Thus sincere instability is not an artifact of harmful or
payoff-neutral intermediate behavior: even a strict better-response path can
lead to an equilibrium that would be blocked under the original preferences.

Proposition~\ref{prop:unstableeq} shows that profitable adjustment does not
recover the sincere stable set.  It leaves a second question: are such dynamics
at least well behaved dynamically, or can profitable tanking cycle?  The key
result is a tanking analogue of the ``Sisterhood'' theorem of Gonczarowski and
Friedgut \cite{GonczarowskiFriedgut2013}.

\begin{restatable}[Tanking Sisterhood / Same-Side Monotonicity]
{theorem}{tankingsisterhood}\label{thm:sisterhood}
Under men-proposing DA with non-tanking men, suppose woman $w$ unilaterally
tanks, producing $Q'\in\mathcal T_w(Q)$. Let
$\mu=\mM(Q)$ and $\eta=\mM(Q')$. If the tanker weakly benefits,
\[
\eta(w)\succeq_w \mu(w),
\]
then every woman weakly benefits:
\[
\eta(v)\succeq_v \mu(v)
\qquad\text{for every }v\in W.
\]
\end{restatable}

The proof is given in Appendix~\ref{app:sisterhood}. It uses the
man-optimality of the pre-deviation DA outcome together with a stable-matching
decomposition argument.

The distinction from standard own-list preference manipulation is important.
Gonczarowski--Friedgut's Sisterhood theorem compares a manipulation with a
fixed sincere baseline.  Under successive own-list manipulations, however, the
currently submitted report may differ from the manipulator's sincere ranking,
so the comparison need not compose from one strategic state to the next.
Under one-sided tanking, the receiving side never changes its own rankings:
$Q_W=P_W$ throughout.  Tanking Sisterhood therefore compares every consecutive
outcome using the same sincere preferences.

\begin{corollary}[Finite strict better-response dynamics]
\label{cor:finite}
Under men-proposing DA with non-tanking men, every maximal sequence of
strictly profitable unilateral tanking actions reaches a tanking equilibrium
after at most $n(n-1)$ actions.
\end{corollary}

The sum of receiving-side sincere partner ranks strictly decreases after every
profitable move and never increases for another receiving-side agent.
Consequently, convergence is not limited to cutoff best responses: \emph{every}
strict better-response path reaches an equilibrium in finitely many
steps.  This contrasts sharply with the standard own-list preference
manipulation model, where strict best-response dynamics can cycle;
Appendix~\ref{app:reportcycle} gives a $3\times3$ example.

Proposition~\ref{prop:unstableeq} and
Corollary~\ref{cor:finite} yield a somewhat unusual conclusion: profitable
tanking dynamics always converge, but the equilibrium they reach need not be
sincerely stable.  Our results characterize the full set of supportable
equilibria and establish convergence of strict better-response dynamics, but
they do not characterize which supportable equilibria are reachable through
such profitable histories.  Understanding that intermediate outcome set---and
how frequently it contains sincerely unstable equilibria---is a natural
equilibrium-selection question for future work.

\section{Conclusion}

Strategic tanking introduces a distinct channel for manipulation in stable matching: an agent can improve their assignment by lowering their position in others' rankings while leaving their own preference ordering unchanged. Our central structural result shows that cutoff tanking produces exactly the same stable-set filter as traditional truncation, despite leaving every pair acceptable and the tanker's own preference list unchanged. This equivalence yields optimal individual and same-side coordinated cutoff responses under broad classes of stable selection rules.

The equilibrium consequences depend sharply on the strategy space. Cutoff tanking remains closely tied to sincere stability: bilateral cutoff equilibria recover exactly the sincere stable set, while one-sided cutoff equilibrium selects the strategic side's sincere-optimal stable matching. Arbitrary tanking can instead reshape the stable set, producing sincerely unstable equilibria and a substantially broader set of supportable outcomes. At the same time, its dynamics retain considerable structure: profitable receiving-side tanking satisfies a state-by-state same-side monotonicity property, and every strict better-response path converges in finitely many steps. Two natural directions remain open: extending the structural and equilibrium theory to general many-to-one markets, and characterizing which supportable tanking equilibria are reachable from the sincere profile through profitable strategic adjustment.

Taken together, these results broaden the usual interpretation of strategic behavior under deferred acceptance. Proposer-side strategyproofness protects agents against manipulating their own reports, and our results show that proposers are also protected against tanking. But this protection does not eliminate strategic behavior from the market. Whenever the sincere stable set contains multiple matchings, profitable receiving-side behavior remains available through both traditional preference manipulation and strategic tanking, even when one side is restricted to honestly reporting its own preferences. Strategyproof preference reporting therefore addresses only one strategic channel when participants can also influence how preferences over them are formed.

\section*{AI-Assistance Disclosure}

OpenAI ChatGPT assisted with drafting, literature discovery, proof development, counterexample searches, and referee-style review. The author supplied the research direction and initial results and independently verified all retained mathematical claims. Detailed information on the AI-assisted development appears in
Appendix~\ref{app:ai}.

\bibliographystyle{unsrtnat}
\bibliography{references}

@article{GaleShapley1962,
 author = {Gale, David and Shapley, Lloyd S.},
 title = {College Admissions and the Stability of Marriage},
 journal = {The American Mathematical Monthly}, year = {1962},
 volume = {69}, number = {1}, pages = {9--15}, doi = {10.2307/2312726}
}

@article{DubinsFreedman1981,
 author = {Dubins, Lester E. and Freedman, David A.},
 title = {Machiavelli and the {Gale--Shapley} Algorithm},
 journal = {The American Mathematical Monthly}, year = {1981},
 volume = {88}, number = {7}, pages = {485--494}, doi = {10.2307/2321753}
}

@article{GaleSotomayor1985,
 author = {Gale, David and Sotomayor, Marilda},
 title = {Ms. {Machiavelli} and the Stable Matching Problem},
 journal = {The American Mathematical Monthly}, year = {1985},
 volume = {92}, number = {4}, pages = {261--268}, doi = {10.2307/2323645}
}

@article{Roth1982,
 author = {Roth, Alvin E.},
 title = {The Economics of Matching: Stability and Incentives},
 journal = {Mathematics of Operations Research}, year = {1982},
 volume = {7}, number = {4}, pages = {617--628}, doi = {10.1287/moor.7.4.617}
}

@article{KowalskiLeary1990,
 author = {Kowalski, Robin M. and Leary, Mark R.},
 title = {Strategic Self-Presentation and the Avoidance of Aversive Events: Antecedents and Consequences of Self-Enhancement and Self-Depreciation},
 journal = {Journal of Experimental Social Psychology}, year = {1990},
 volume = {26}, number = {4}, pages = {322--336}, doi = {10.1016/0022-1031(90)90042-K}
}

@article{BeckerMartin1995,
 author = {Becker, Thomas E. and Martin, Scott L.},
 title = {Trying to Look Bad at Work: Methods and Motives for Managing Poor Impressions in Organizations},
 journal = {Academy of Management Journal}, year = {1995},
 volume = {38}, number = {1}, pages = {174--199}, doi = {10.5465/256732}
}

@article{GibsonSachau2000,
 author = {Gibson, Bryan and Sachau, Daniel},
 title = {Sandbagging as a Self-Presentational Strategy: Claiming to Be Less than You Are},
 journal = {Personality and Social Psychology Bulletin}, year = {2000},
 volume = {26}, number = {1}, pages = {56--70}, doi = {10.1177/0146167200261006}
}

@article{McLuhan2020,
 author = {McLuhan, Arthur},
 title = {Adopting a Cloak of Incompetence: Impression Management Techniques for Feigning Lesser Selves},
 journal = {Sociological Theory}, year = {2020},
 volume = {38}, number = {2}, pages = {122--141}, doi = {10.1177/0735275120926222}
}

@article{WinchesterEtAl2026,
 author = {Winchester, Christopher C. and Hsu, Emily and Campbell, Elizabeth M. and Call, Matthew L.},
 title = {Underperformance by Design: A Scoping Review and Research Agenda of Intentional Task Underperformance at Work},
 journal = {Journal of Management}, year = {2026},
 volume = {52}, number = {6}, pages = {2395--2426}, doi = {10.1177/01492063251398774},
 note = {First published online December 26, 2025}
}

@article{CharnessKuhnVilleval2011,
 author = {Charness, Gary and Kuhn, Peter and Villeval, Marie Claire},
 title = {Competition and the Ratchet Effect},
 journal = {Journal of Labor Economics}, year = {2011},
 volume = {29}, number = {3}, pages = {513--547}, doi = {10.1086/659347}
}

@article{TeoSethuramanTan2001,
 author = {Teo, Chung-Piaw and Sethuraman, Jay and Tan, Wee-Peng},
 title = {{Gale--Shapley} Stable Marriage Problem Revisited: Strategic Issues and Applications},
 journal = {Management Science}, year = {2001},
 volume = {47}, number = {9}, pages = {1252--1267}, doi = {10.1287/mnsc.47.9.1252.9784}
}

@article{ShenDengTang2021,
 author = {Shen, Weiran and Deng, Yuan and Tang, Pingzhong},
 title = {Coalitional Permutation Manipulations in the {Gale--Shapley} Algorithm},
 journal = {Artificial Intelligence}, year = {2021},
 volume = {301}, pages = {103577}, doi = {10.1016/j.artint.2021.103577}
}

@inproceedings{HosseiniUmarVaish2021,
 author = {Hosseini, Hadi and Umar, Fatima and Vaish, Rohit},
 title = {Accomplice Manipulation of the Deferred Acceptance Algorithm},
 booktitle = {Proceedings of the Thirtieth International Joint Conference on Artificial Intelligence (IJCAI-21)},
 year = {2021}, pages = {231--237}, publisher = {International Joint Conferences on Artificial Intelligence Organization},
 doi = {10.24963/ijcai.2021/33}
}

@misc{HosseiniLisowskiPathak2025,
 author = {Hosseini, Hadi and Lisowski, Grzegorz and Pathak, Shraddha},
 title = {Two-Sided Manipulation Games in Stable Matching Markets},
 year = {2025}, eprint = {2506.00554}, archivePrefix = {arXiv},
 doi = {10.48550/arXiv.2506.00554}, note = {Preprint}
}

@inproceedings{RastegariEtAl2013,
 author = {Rastegari, Baharak and Condon, Anne and Immorlica, Nicole and Leyton-Brown, Kevin},
 title = {Two-Sided Matching with Partial Information},
 booktitle = {Proceedings of the Fourteenth ACM Conference on Electronic Commerce (EC '13)},
 publisher = {Association for Computing Machinery}, year = {2013},
 pages = {733--750}, doi = {10.1145/2482540.2482607}
}

@inproceedings{RastegariGoldbergManlove2016,
 author = {Rastegari, Baharak and Goldberg, Paul and Manlove, David},
 title = {Preference Elicitation in Matching Markets via Interviews: A Study of Offline Benchmarks},
 booktitle = {Proceedings of the 15th International Conference on Autonomous Agents and Multiagent Systems (AAMAS 2016)},
 publisher = {International Foundation for Autonomous Agents and Multiagent Systems},
 year = {2016}, pages = {1393--1394}, note = {Extended abstract; full version arXiv:1602.04792}
}

@inproceedings{BaileyTovey2025,
 author = {Bailey, James P. and Tovey, Craig A.},
 title = {On the {Gale--Shapley} Algorithm for Stable Matchings with a Partial Honesty {Nash} Refinement},
 booktitle = {Proceedings of the 24th International Conference on Autonomous Agents and Multiagent Systems (AAMAS 2025)},
 publisher = {International Foundation for Autonomous Agents and Multiagent Systems},
 year = {2025}, pages = {197--204},
 url = {https://www.ifaamas.org/Proceedings/aamas2025/pdfs/p197.pdf}
}

@article{AbdulkadirogluSonmez2003,
 author = {Abdulkadiro{\u{g}}lu, Atila and S{\"o}nmez, Tayfun},
 title = {School Choice: A Mechanism Design Approach},
 journal = {American Economic Review}, year = {2003},
 volume = {93}, number = {3}, pages = {729--747}, doi = {10.1257/000282803322157061}
}

@article{BalinskiSonmez1999,
  author = {Balinski, Michel and S{\"o}nmez, Tayfun},
  title = {A Tale of Two Mechanisms: Student Placement},
  journal = {Journal of Economic Theory},
  year = {1999},
  volume = {84},
  number = {1},
  pages = {73--94}
}

@article{HatfieldKojimaNarita2016,
  author = {Hatfield, John William and Kojima, Fuhito and Narita, Yusuke},
  title = {Improving Schools through School Choice: A Market Design Approach},
  journal = {Journal of Economic Theory},
  year = {2016},
  volume = {166},
  pages = {186--211},
  doi = {10.1016/j.jet.2016.07.001}
}

@article{KlijnPaisVorsatz2020,
  author = {Klijn, Flip and Pais, Joana and Vorsatz, Marc},
  title = {Improving Schools through School Choice: An Experimental Study of Deferred Acceptance},
  journal = {Economics Letters},
  year = {2020},
  volume = {186},
  pages = {108853},
  doi = {10.1016/j.econlet.2019.108853}
}

@article{KobayashiMatsui2009,
  author = {Kobayashi, Hirotatsu and Matsui, Tomomi},
  title = {Successful Manipulation in Stable Marriage Model with Complete Preference Lists},
  journal = {IEICE Transactions on Information and Systems},
  year = {2009},
  volume = {E92-D},
  number = {2},
  pages = {116--119},
  doi = {10.1587/transinf.E92.D.116}
}

@article{KobayashiMatsui2010,
  author = {Kobayashi, Hirotatsu and Matsui, Tomomi},
  title = {Cheating Strategies for the {Gale--Shapley} Algorithm with Complete Preference Lists},
  journal = {Algorithmica},
  year = {2010},
  volume = {58},
  number = {1},
  pages = {151--169}
}

@article{GonczarowskiFriedgut2013,
  author = {Gonczarowski, Yannai A. and Friedgut, Ehud},
  title = {Sisterhood in the {Gale--Shapley} Matching Algorithm},
  journal = {The Electronic Journal of Combinatorics},
  year = {2013},
  volume = {20},
  number = {2},
  pages = {P12},
  doi = {10.37236/3267}
}

@inproceedings{AthanasopoulosGeorgeDimitrakakis2025,
  author = {Athanasopoulos, Andreas and George, Anne-Marie and Dimitrakakis, Christos},
  title = {Probably Correct Optimal Stable Matching for Two-Sided Markets Under Uncertainty},
  booktitle = {Proceedings of the 24th International Conference on Autonomous Agents and Multiagent Systems (AAMAS 2025)},
  year = {2025},
  pages = {152--160},
  publisher = {International Foundation for Autonomous Agents and Multiagent Systems},
  url = {https://www.ifaamas.org/Proceedings/aamas2025/pdfs/p152.pdf}
}

@inproceedings{HosseiniRoyZhang2024,
  author = {Hosseini, Hadi and Roy, Sanjukta and Zhang, Duohan},
  title = {Putting {Gale \& Shapley} to Work: Guaranteeing Stability Through Learning},
  booktitle = {Advances in Neural Information Processing Systems},
  year = {2024},
  volume = {37},
  url = {https://proceedings.neurips.cc/paper_files/paper/2024/file/7fc0aac0d700873ac520af9b05e3f2b6-Paper-Conference.pdf}
}

@inproceedings{AshlagiEtAl2025,
  author = {Ashlagi, Itai and Chen, Jiale and Roghani, Mohammad and Saberi, Amin},
  title = {Stable Matching with Interviews},
  booktitle = {16th Innovations in Theoretical Computer Science Conference (ITCS 2025)},
  series = {Leibniz International Proceedings in Informatics (LIPIcs)},
  volume = {325},
  year = {2025},
  pages = {12:1--12:19},
  doi = {10.4230/LIPIcs.ITCS.2025.12}
}

@inproceedings{HosseiniUmarVaish2022,
  author = {Hosseini, Hadi and Umar, Fatima and Vaish, Rohit},
  title = {Two for One \& One for All: Two-Sided Manipulation in Matching Markets},
  booktitle = {Proceedings of the Thirty-First International Joint Conference on Artificial Intelligence (IJCAI-22)},
  year = {2022},
  pages = {321--327},
  doi = {10.24963/ijcai.2022/46}
}

@misc{Gonczarowski2014Blacklists,
  author = {Gonczarowski, Yannai A.},
  title = {Manipulation of Stable Matchings Using Minimal Blacklists},
  year = {2014},
  eprint = {1307.7477},
  archivePrefix = {arXiv},
  doi = {10.48550/arXiv.1307.7477},
  note = {Full preprint version, revised March 10, 2014}
}

@article{ConnellyEtAl2012,
  author = {Connelly, Catherine E. and Zweig, David and Webster, Jane and Trougakos, John P.},
  title = {Knowledge Hiding in Organizations},
  journal = {Journal of Organizational Behavior},
  year = {2012},
  volume = {33},
  number = {1},
  pages = {64--88},
  doi = {10.1002/job.737}
}

@article{ShepperdSocherman1997,
  author = {Shepperd, James A. and Socherman, Robert E.},
  title = {On the Manipulative Behavior of Low Machiavellians: Feigning Incompetence to ``Sandbag'' an Opponent},
  journal = {Journal of Personality and Social Psychology},
  year = {1997},
  volume = {72},
  number = {6},
  pages = {1448--1459},
  doi = {10.1037/0022-3514.72.6.1448}
}

@article{SachauEtAl2014,
  author = {Sachau, Daniel and Simmering, Luke and Adler, Max and Ryan, Warren},
  title = {Sandbagging: Faking Incompetence on the Golf Course},
  journal = {International Journal of Golf Science},
  year = {2014},
  volume = {3},
  number = {1},
  pages = {64--77},
  doi = {10.1123/ijgs.2013-0001}
}

@article{ProdanBoza2025,
  author = {Prodan, Livioara E. and Boza, Mihaela},
  title = {Exploring Sandbagging Behavior: A Systematic Review},
  journal = {Journal of Individual Differences},
  year = {2025},
  volume = {46},
  number = {2},
  pages = {109--119},
  doi = {10.1027/1614-0001/a000436}
}

@article{Stadnicka2024,
  author = {Stadnicka, Agnieszka},
  title = {Feigned Digital Incompetence as a New Managerial Challenge: A Systematic Literature Review and Future Research Agenda},
  journal = {Annales Universitatis Mariae Curie-Sk{\l}odowska, sectio H -- Oeconomia},
  year = {2024},
  volume = {58},
  number = {4},
  pages = {223--248},
  doi = {10.17951/h.2024.58.4.223-248}
}

@techreport{NRMPApplicantSurvey2024,
  author = {{National Resident Matching Program}},
  title = {Charting Outcomes: Applicant Survey Results, Main Residency Match, 2024: Introduction and Methods},
  institution = {National Resident Matching Program},
  year = {2024},
  url = {https://www.nrmp.org/wp-content/uploads/2024/09/Intro-and-Methods_2024-Applicant-Survey-Report-08232024.pdf}
}

@misc{NRMPAlgorithm,
  author = {{National Resident Matching Program}},
  title = {How the Matching Algorithm Works},
  howpublished = {Official NRMP explanation},
  url = {https://www.nrmp.org/intro-to-the-match/how-matching-algorithm-works/},
  note = {Accessed September 19, 2026}
}

@inproceedings{AthanasopoulosDimitrakakis2026,
  author    = {Andreas Athanasopoulos and Christos Dimitrakakis},
  title     = {Learning in Matching Games with Bandit Feedback},
  booktitle = {Joint European Conference on Machine Learning and Knowledge Discovery in Databases (ECML PKDD)},
  pages     = {687--702},
  publisher = {Springer},
  year      = {2026}
}

\newpage

\appendix
\section{Supporting Proofs}\label{app:proofs}

\subsection{The local stable-set lemma}\label{app:local}
\localstablelemma*
\begin{proof}
Suppose $i=w\in W$ moves downward in man $m$'s list; the other case is symmetric. If $\mu$ is stable before the move but not after it, a newly blocking pair must involve a comparison that changed. Only comparisons between $w$ and members of $X$ changed, and every crossed woman moved upward relative to $w$. Therefore, a new blocker can arise only against a matching assigning $m$ to $w$, proving the first assertion.

For the second assertion, reverse the move one crossed element at a time, moving each element of $X$ back below $w$ until the original list is restored. If $\mu$ is stable after the full demotion but not before it, stability is lost at some first reverse step. Applying the first assertion to that single reverse move shows that $m$ is matched to the woman moved across $w$ at that step, hence to an element of $X$.
\end{proof}

\subsection{Exact cutoff filtering and truncation}\label{app:filter}
\exactcutofffilter*
\begin{proof}
It is enough to prove the case $i=w\in W$; the other side is symmetric. Let $\mu\in\SM(Q)$ assign $w$ at least $t$. She tanks only at men she ranks below $t$, none of whom is $\mu(w)$. Lemma~\ref{lem:local}, applied one list at a time, preserves $\mu$. In particular, a stable matching $\nu$ with $\nu(w)=t$ survives.

Let $\lambda=\mM(Q^{w,t})$. If $\lambda(w)=m$ with $t\succ_w^Q m$, then $m$ ranks $w$ last after the cutoff, while $\nu(m)\ne w$. Hence $m$ strictly prefers $\nu(m)$ to $w$, contradicting man-optimality of $\lambda$. Thus $\lambda(w)\succeq_w^Q t$, and woman-pessimality implies that every induced stable matching satisfies the cutoff. Finally, restoring the crossed comparisons cannot create a blocker involving $w$, because every restored man lies below $t$ while her induced-stable partner lies at least $t$. Therefore every induced stable matching was already stable at $Q$.
\end{proof}

\truncationequivalence*
\begin{proof}
Again take $i=w\in W$. Theorem~\ref{thm:filter} gives the tanking equality. Let $R$ be the profile obtained by truncating $w$'s own list below stable partner $t$. A stable matching $\nu$ with $\nu(w)=t$ remains stable at $R$, so $w$ is matched in at least one stable matching of the truncated instance. By invariance of the unmatched set across stable matchings \cite[Proposition~1]{GaleSotomayor1985}, she is matched in every stable matching of $R$, necessarily to a listed partner weakly above $t$.

Every stable matching of $Q$ assigning $w$ at least $t$ remains stable after truncation because only potential blocking pairs are removed. Conversely, let $\mu$ be stable at $R$. Restoring the deleted partners cannot create a blocker $(m,w)$: every restored man is below $t$, whereas $\mu(w)$ is at least $t$. Thus $\mu$ is stable at $Q$ and satisfies the cutoff. The symmetric proof applies to a man.
\end{proof}

\jointcutofffilter*
\begin{proof}
Assume $C\subseteq W$; the other side is symmetric. Because members of $C$ tank only in men's rankings, their own rankings and one another's cutoff comparisons do not change. Apply Theorem~\ref{thm:filter} sequentially to the coalition members. Joint feasibility guarantees that a common stable matching satisfying every cutoff survives every step, so each later target remains a stable partner. After all actions, the surviving set is exactly the intersection of the individual cutoff conditions. Because intersection is order independent, the final stable set is independent of the coalition's action order.
\end{proof}

\subsection{Unilateral and coordinated upper bounds}\label{app:upper}
\unilateralupper*
\begin{proof}
It is enough to consider $i=w\in W$. Let $Q'\in\mathcal T_w(Q)$ and $\eta\in\SM(Q')$, and write $t=\eta(w)$. Restore all men's rankings to their values in $Q$. In an auxiliary profile, additionally move $t$ to the top of $w$'s own list, preserving the order of all other men.

The matching $\eta$ is stable in this auxiliary profile. Restoring $w$ upward in men's lists can create a new blocking pair only involving $w$; but $w$ now ranks her assigned partner $t$ first, so no such pair blocks. Women-proposing DA in the auxiliary profile therefore assigns $w$ the partner $t$: the DA outcome is woman-optimal and $t$ is already her first choice among a stable matching.

Relative to the original current profile $Q$, the auxiliary change is solely an own-list misreport by proposer $w$. Proposer strategyproofness therefore implies that $t$ cannot be strictly better under $Q_w$ than $w$'s woman-optimal stable partner at $Q$. This proves the bound. Exchanging the two sides proves the symmetric statement.
\end{proof}

\coalitionalupper*
\begin{proof}
The proof follows identically to the single-agent case: 
Assume $C\subseteq W$; the other side is symmetric. Suppose coordinated tanking by $C$ produces a profile $Q'$ and a stable matching $\eta\in\SM(Q')$ with $\eta(w)\succ_w^Q b_w(Q)$ for every $w\in C$. Restore the men's rankings to $Q_M$. For each $w\in C$, form an auxiliary own-list report that places $\eta(w)$ first and otherwise preserves the current order.

The matching $\eta$ is stable in this auxiliary profile. Restoring a coalition woman upward in men's lists can create a new blocking pair only involving that woman, but every coalition woman now ranks her assigned partner first. Under women-proposing DA, each member of $C$ therefore receives $\eta(w)$: because $\eta$ is stable and $\eta(w)$ is first in her reported list, the woman-optimal stable outcome cannot assign her anything else. Relative to $Q$, this is an own-list misreport by a coalition of proposers that makes every coalition member strictly better than her truthful woman-optimal partner $b_w(Q)$, contradicting Dubins--Freedman's coalitional proposer result \cite{DubinsFreedman1981}. The symmetric proof applies to a coalition in $M$.
\end{proof}

\subsection{Cutoff equilibria}\label{app:cutoffeq}
\survivingorder*
\begin{proof}
Consider one cutoff step by agent $j$. If the action changes the relative order of $j$ and another candidate in agent $i$'s list, then $j$ has tanked at $i$. Because $i$ lies below $j$'s cutoff target in $j$'s current ranking, Theorem~\ref{thm:filter} removes every stable matching assigning $j$ to $i$. Thus no partner whose position relative to another surviving stable partner is changed by this step remains attainable for the affected list owner. Applying this argument sequentially along the cutoff history shows that the original and current rankings agree on the relative order of all partners appearing in $\SM(Q)$.
\end{proof}

\bilateralcutoffeq*
\begin{proof}
Let the stable selection rule choose $\mu\in\SM(Q)$. If $\SM(Q)$ is non-singleton, then $\mM(Q)\ne\mW(Q)$. If $\mu\ne\mM(Q)$, some man strictly prefers his partner in $\mM(Q)$ to $\mu(m)$ under $Q_m$; if $\mu=\mM(Q)$, then $\mu\ne\mW(Q)$ and some woman strictly prefers her partner in $\mW(Q)$. By Lemma~\ref{lem:survivingorder}, the same strict preference holds under the sincere ranking $P_i$. Corollary~\ref{cor:bestresponse} therefore gives that agent a profitable cutoff under either equilibrium interpretation.

Conversely, if $\SM(Q)$ is a singleton, every agent's best current stable partner is already its selected partner. A cutoff can only filter the stable set further and cannot expose a better stable partner, so no cutoff deviation is profitable under either interpretation.
\end{proof}

\cutoffsupport*
\begin{proof}
Every current-state cutoff satisfies $\SM(Q^{k+1})\subseteq\SM(Q^k)$ by Theorem~\ref{thm:filter}. Hence every stable outcome reached by a cutoff history from $P$ lies in $\SM(P)$, giving one inclusion.

For the reverse inclusion, fix $\mu\in\SM(P)$. First let every woman apply a cutoff at $\mu(w)$. Theorem~\ref{thm:coalitionfilter} leaves precisely the sincere stable matchings $\nu$ for which every woman weakly prefers $\nu$ to $\mu$; in particular, $\mu$ survives. By the stable lattice, every man weakly prefers $\mu$ to each such survivor. Lemma~\ref{lem:survivingorder} implies that these comparisons are preserved in the current induced rankings. Now let every man cutoff at $\mu(m)$. The target $\mu$ survives, while every other surviving matching differs from $\mu$ at some man who strictly prefers his $\mu$-partner and is therefore eliminated. The final stable set is $\{\mu\}$, which is a cutoff equilibrium under either interpretation by Corollary~\ref{cor:cutoffsingleton}. Thus every sincere stable matching is supportable.
\end{proof}

\onesidecutoff*
\begin{proof}
Assume the women are the only tankers. Their own rankings never change. We first show that $\mW(P)$ survives every cutoff history. Inductively, suppose it survives at the current state $Q$. Any woman's cutoff target $t$ is a current stable partner. Because every current stable matching is sincerely stable by stable-set contraction and $\mW(P)$ is the sincere best stable matching for all women, $\mW(P)(w)\succeq_w t$. Theorem~\ref{thm:filter} therefore preserves $\mW(P)$.

If a stable selection rule at $Q$ returns $\mu\ne\mW(P)$, some woman strictly prefers $\mW(P)(w)$ to $\mu(w)$. Since $\mW(P)$ still lies in $\SM(Q)$, that partner remains a current stable partner, and Corollary~\ref{cor:bestresponse} gives a profitable cutoff. Hence no such outcome is an equilibrium. Conversely, the jointly feasible cutoffs $t_w=\mW(P)(w)$ leave only $\mW(P)$ by Theorem~\ref{thm:coalitionfilter}, giving an equilibrium profile. The other side is symmetric.
\end{proof}

\subsection{Arbitrary tanking under DA}\label{app:arbitrary}
\unstableeq*
The verification is given in Appendix~\ref{app:unstable}: the action in Example~\ref{ex:unstable} is strictly profitable, the induced stable set is a singleton, and the resulting matching has a sincere blocking pair.

\uniqueinduced*
\begin{proof}
If $\SM(Q)$ is non-singleton, then $\mM(Q)\ne\mW(Q)$ and some woman strictly prefers $\mW(Q)(w)$ to $\mM(Q)(w)$. Corollary~\ref{cor:bestresponse} gives a cutoff tanking action that forces her induced woman-optimal partner under every stable selection rule, including men-proposing DA. Thus $Q$ is not a tanking equilibrium.

If $\SM(Q)$ is a singleton, every woman's best induced stable partner is her current DA partner. Theorem~\ref{thm:upper}, applied locally at $Q$, says that no further self-demotion can produce any induced-stable outcome she prefers. Hence no woman has a profitable further tanking deviation.
\end{proof}

\subsection{Tanking Sisterhood}\label{app:sisterhood}
\tankingsisterhood*

\begin{proof}
Suppose, toward a
contradiction, that some woman is worse off, and define
\[
B=\{v\in W:\mu(v)\succ_v \eta(v)\}.
\]
as the set of women that are worse off. 
By hypothesis, $w\notin B$. Let $A=\mu(B)$
be the set of men matched under $\mu$ to women in $B$.

We first show that $\eta(A)=B$.
Fix $v\in B$ and let $m=\mu(v)\in A$. Since
$v$ strictly prefers $m$ to $\eta(v)$, stability of $\eta$ at $Q'$
implies
\[
\eta(m)\succ^{Q'}_m v.
\]
This comparison also holds at $Q$. Indeed, if $\eta(m)\neq w$, neither
side of the comparison was changed by $w$'s tanking. If
$\eta(m)=w$, then $w$ was moved only downward in $m$'s ranking, so
\[
w\succ^{Q'}_m v
\quad\Longrightarrow\quad
w\succ^{Q}_m v.
\]
Hence, in either case,
\[
\eta(m)\succ^{Q}_m v=\mu(m).
\]

Now use stability of $\mu$ at $Q$. Since $m$ strictly prefers
$\eta(m)$ to $\mu(m)$, the pair $(m,\eta(m))$ can fail to block
$\mu$ only if
\[
\mu(\eta(m))\succ_{\eta(m)} m.
\]
But $m=\eta(\eta(m))$, so this says precisely that
$\eta(m)\in B$. Thus every $m\in A$ is matched by $\eta$ to a woman
in $B$, and therefore
\[
\eta(A)\subseteq B.
\]
Because $\eta$ is a bijection and $|A|=|B|$, we conclude that
\[
\eta(A)=B.
\]

We can therefore define a perfect matching $\widehat{\mu}$ by
splicing $\eta$ on $A\cup B$ with $\mu$ elsewhere:
\[
\widehat{\mu}(m)=
\begin{cases}
\eta(m), & m\in A,\\
\mu(m), & m\notin A.
\end{cases}
\]
Equivalently, women in $B$ receive their $\eta$-partners and women
outside $B$ receive their $\mu$-partners. We claim that
$\widehat{\mu}$ is stable at $Q$.

Consider any pair $(m,v)$.

\emph{Case 1: $m\notin A$ and $v\notin B$.}
Both agents have their $\mu$-partners under $\widehat{\mu}$, so a
blocking pair would also block $\mu$, contradicting stability of
$\mu$ at $Q$.

\emph{Case 2: $m\in A$ and $v\in B$.}
Both agents have their $\eta$-partners under $\widehat{\mu}$. Since
$\eta(A)=B$ and $w\notin B$, neither $v$ nor $\eta(m)$ is the
tanker $w$. Thus $m$'s comparison between $v$ and $\eta(m)$ is
unchanged from $Q$ to $Q'$, while women's rankings do not change at
all. Hence any blocking pair for $\widehat{\mu}$ at $Q$ would also
block $\eta$ at $Q'$, contradicting stability of $\eta$.

\emph{Case 3: $m\in A$ and $v\notin B$.}
From the argument above, every $m\in A$ satisfies
\[
\eta(m)\succ^Q_m \mu(m).
\]
If $m$ preferred $v$ to $\widehat{\mu}(m)=\eta(m)$, then
\[
v\succ^Q_m \eta(m)\succ^Q_m\mu(m).
\]
Stability of $\mu$ would therefore imply
\[
\mu(v)\succ_v m.
\]
Since $v\notin B$ receives $\mu(v)$ under $\widehat{\mu}$, the pair
$(m,v)$ cannot block.

\emph{Case 4: $m\notin A$ and $v\in B$.}
Suppose, for contradiction, that $(m,v)$ blocks
$\widehat{\mu}$ at $Q$. Then
\[
m\succ_v\eta(v)
\qquad\text{and}\qquad
v\succ^Q_m\mu(m).
\]
Women's rankings are unchanged, so $v$ also prefers $m$ to
$\eta(v)$ at $Q'$. Stability of $\eta$ at $Q'$ therefore implies
\[
\eta(m)\succ^{Q'}_m v.
\]
As above, this comparison also holds at $Q$: if
$\eta(m)\neq w$ it is unchanged, while if $\eta(m)=w$, tanking only
moved $w$ downward. Consequently,
\[
\eta(m)\succ^Q_m v\succ^Q_m\mu(m).
\]
Stability of $\mu$ at $Q$ then requires
\[
\mu(\eta(m))\succ_{\eta(m)} m.
\]
Since $m=\eta(\eta(m))$, this implies $\eta(m)\in B$. But
$m\notin A$ and $\eta(A)=B$, so $\eta(m)\notin B$, a contradiction.
Thus $\widehat{\mu}$ is stable at $Q$.

Finally, for every $m\in A$ we established
\[
\widehat{\mu}(m)=\eta(m)\succ^Q_m\mu(m).
\]
Since $B$ was assumed nonempty, $A$ is nonempty. Hence
$\widehat{\mu}$ is a stable matching at $Q$ that gives at least one
man a partner strictly better than his partner under
$\mu=\mu_M(Q)$, contradicting the man-optimality of $\mu$.
Therefore $B=\varnothing$, and every woman weakly benefits.
\end{proof}




\subsection{Complete-list implementation and arbitrary-tanking equilibria}\label{app:support}
We first record the reachability observation used by the support theorem.

\begin{lemma}[Reachability of men's rankings]\label{lem:reach}
From any baseline men's profile $P_M$, women can generate any strict complete target profile $R_M$ through a finite sequence of self-demotions.
\end{lemma}
\begin{proof}
Fix one man's desired final ranking $v_1\succ v_2\succ\cdots\succ v_n$. Starting from his current list, have $v_2$ move to the bottom, then $v_3$, and continue through $v_n$, skipping a move that changes nothing. The resulting order is exactly the target order. Repeat independently for every man.
\end{proof}

\supporttheorem*
\begin{proof}
By Theorem~\ref{thm:equilibriumunique}, a tanking equilibrium $Q$ has a unique stable matching $\mu$. Thus both orientations of DA return $\mu$, and $\mu=\mW(Q_M,P_W)$. This gives one inclusion.

Conversely, suppose $\mu=\mW(R_M,P_W)$ for some men's profile $R_M$. Promote each man's final partner $\mu(m)$ to the top of his list, preserving every other comparison; call the resulting profile $Q_M$. The women-proposing DA run that produced $\mu$ under $R_M$ is unchanged: whenever a man receives a proposal from his final partner, that partner already outranked every other woman who had proposed to him, so moving her to the top cannot alter any rejection that led to the final outcome. Thus women-proposing DA still returns $\mu$.

Under $Q_M$, every man ranks his distinct $\mu$-partner first, so men-proposing DA returns $\mu$ immediately. The two extremes coincide; therefore $\mu$ is the unique stable matching at $(Q_M,P_W)$. Lemma~\ref{lem:reach} shows that women can reach $Q_M$ from any baseline $P_M$ by tanking, and Theorem~\ref{thm:equilibriumunique} makes the resulting profile a tanking equilibrium.

It remains to show that membership in the set on the right-hand side can be
tested, and that a supporting tanking-equilibrium profile can be constructed,
in $O(n^2)$ time.

Kobayashi and Matsui study exactly the corresponding complete-list
implementation problem \cite{KobayashiMatsui2009,KobayashiMatsui2010}.
Given the complete preference lists of one side and a target matching $\mu$,
they characterize when complete preference lists for the opposite side can be
chosen so that proposer-deferred acceptance returns $\mu$. Their main theorem
gives a necessary and sufficient condition based on the strongly connected
components of an associated directed graph and yields an $O(n^2)$ algorithm
for both testing implementability and constructing the required preference
lists.

Applying their result with the roles of men and women reversed, we can test in
$O(n^2)$ time whether there exists a men's profile $R_M$ such that $\mu=\mu_W(R_M,P_W)$,
and, whenever one exists, construct such a profile $R_M$.  The construction
above then converts this profile into a uniquely stable profile supporting
$\mu$ as a tanking equilibrium, also in $O(n^2)$ time.
\end{proof}

\subsection{Pareto-efficient targets}\label{app:pareto}
\paretocor*
\begin{proof}
Let $\mu$ be a perfect matching that is Pareto efficient for the receiving
side. Construct a proposing-side profile $R_M$ by ranking $\mu(m)$ first in
each proposer $m$'s list, with the remaining agents ordered arbitrarily.
Proposer-deferred acceptance then returns $\mu$, so $\mu$ is the
proposer-optimal stable matching of $(R_M,P_W)$.

Suppose there were another stable matching $\nu\neq\mu$. Since $\mu$ is
proposer-optimal, it is receiver-pessimal among the stable matchings.
Therefore every receiving-side agent weakly prefers $\nu$ to $\mu$, and at
least one strictly prefers $\nu$. This contradicts the Pareto efficiency of
$\mu$ with respect to $P_W$.

Hence $\mu$ is the unique stable matching of $(R_M,P_W)$ and, in particular,  $\mu=\mu_W(R_M,P_W)$.
Theorem~\ref{thm:support} therefore implies that $\mu$ is supportable at a
tanking equilibrium.

The reverse inclusion fails by the example in in
Appendix~\ref{app:inefficient}.
\end{proof}

\section{Comparative Examples}\label{app:examples}
\subsection{The sincerely unstable tanking equilibrium}\label{app:unstable}
After the action in Example~\ref{ex:unstable}, the men's lists are
\[
\begin{aligned}
 m_1&:w_3,w_2,w_1, & m_2&:w_1,w_2,w_3,\\
 m_3&:w_1,w_3,w_2.&&
\end{aligned}
\]
Men-proposing DA returns
\[
(m_1,w_2),(m_2,w_1),(m_3,w_3).
\]
Under women-proposing DA, $w_1$ first proposes to $m_1$, while $w_2,w_3$ propose to $m_3$; continuing rejections leads to the same matching. Thus the induced stable set is a singleton and, by Theorem~\ref{thm:equilibriumunique}, the profile is a tanking equilibrium. Sincerely, $w_2$ prefers $m_3$ to $m_1$ and $m_3$ prefers $w_2$ to $w_3$, so $(m_3,w_2)$ blocks. The move is strictly profitable for $w_2$, so the equilibrium is reached from sincerity in one strict better response.

\subsection{A best-response cycle under traditional own-list preference manipulation}\label{app:reportcycle}
Corollary~\ref{cor:finite} relies on the strategic side's own rankings remaining sincere. Standard own-list preference manipulation does not share this finite-improvement property. Consider fixed men's rankings
\[
 m_1:w_1,w_2,w_3,\qquad
 m_2:w_1,w_2,w_3,\qquad
 m_3:w_2,w_1,w_3,
\]
and sincere women's rankings
\[
 w_1:m_3,m_1,m_2,\qquad
 w_2:m_3,m_1,m_2,\qquad
 w_3:m_1,m_2,m_3.
\]
Keep $w_3$ truthful. Under men-proposing DA, the following complete reports and sincere outcomes form a cycle:
\[
\begin{array}{c|c|c|c}
& w_1\text{ reports} & w_2\text{ reports} & (\mu(w_1),\mu(w_2),\mu(w_3))\\
\hline
A&m_3,m_1,m_2&m_1,m_3,m_2&(m_1,m_3,m_2)\\
B&m_3,m_2,m_1&m_1,m_3,m_2&(m_3,m_1,m_2)\\
C&m_3,m_2,m_1&m_2,m_3,m_1&(m_2,m_3,m_1)\\
D&m_3,m_1,m_2&m_2,m_3,m_1&(m_3,m_2,m_1)\\
A&m_3,m_1,m_2&m_1,m_3,m_2&(m_1,m_3,m_2).
\end{array}
\]
The deviator at each transition is, respectively, $w_1,w_2,w_1,w_2$, and in every case she moves to $m_3$, her sincere first choice. Thus every deviation is a strict best response, yet the sequence cycles. This does not contradict Gonczarowski--Friedgut: their sisterhood comparison is anchored to a fixed truthful baseline, while the intermediate reports above are themselves strategic.

\subsection{An inefficient arbitrary-tanking equilibrium}\label{app:inefficient}
Consider
\[
\begin{array}{c@{\qquad}c}
 m_1:w_1,w_3,w_2 & w_1:m_2,m_1,m_3\\
 m_2:w_2,w_3,w_1 & w_2:m_1,m_2,m_3\\
 m_3:w_3,w_1,w_2 & w_3:m_1,m_3,m_2.
\end{array}
\]
The diagonal matching $\mu=(m_1,w_1),(m_2,w_2),(m_3,w_3)$ is uniquely stable and therefore a tanking equilibrium, yet exchanging the partners of $w_1$ and $w_2$ resulting in $\nu =(m_1,w_2),(m_2,w_1),(m_3,w_3)$ strictly benefits both women. Thus, an arbitrary tanking equilibrium need not be woman-Pareto efficient.

As established in Corollary \ref{cor:pareto}, this woman-Pareto efficient outcome can be obtained at an equilibrium. If $w_3$ moves below $w_2$ at $m_1$, the DA outcome remains diagonal, but the matching $\nu$ becomes stable. Woman $w_1$ can then tank at $m_1$ and obtain $m_2$, while $w_2$ obtains $m_1$ and helper $w_3$ remains at $m_3$. 
The resulting profile has a unique stable matching and therefore is a tanking equilibrium.
This separates all-strict coalition improvement from coalitions containing indifferent helpers.

\section{AI-Assistance Record}\label{app:ai}

The motivating question for this paper predates the use of generative AI in
its development. The initial motivation arose from an AAMAS 2025 talk by
Andreas Athanasopoulos on stable matching when agents' preferences are
uncertain and gradually learned through noisy feedback. That work prompted our question on  how such preferences are formed and whether an agent could
strategically manipulate the process by which other agents learn or construct
their preferences.

We subsequently developed initial notes on what became strategic
tanking. These notes proposed studying the new strategic action by following
the classical progression of strategic questions in stable matching, identified
the close connection between tanking and traditional truncation, and contained
partial proofs and several of the paper's main structural results. A local
stable-set lemma from earlier work on strategic stable matching
was particularly useful in developing these arguments. The project was then
set aside for more than a year because of other responsibilities.

When the project was resumed, we used OpenAI ChatGPT as a research
assistant, conceptually much like a junior graduate student working from an
existing research direction. The author supplied the existing notes, partial
proofs, preliminary results, and intended organization and asked the model to
construct a full manuscript draft. The draft was subsequently revised through
approximately six major iterations. In each iteration, we reviewed the
previous version and directed changes to proofs, theorem statements, related
work, attribution, scope, examples, and the overall narrative. OpenAI ChatGPT,
including GPT-5.6 Sol during the final revision process, also assisted with
literature discovery, LaTeX preparation, counterexample searches, exploratory
computational checks, and prose revision. Exact model-version information was
not retained for every earlier interaction.

Two substantive results retained in the final paper arose during this
AI-assisted development rather than appearing in our original notes.
First, the model identified the same-side monotonicity phenomenon developed in
Theorem~\ref{thm:sisterhood}. It also identified the connection to the
classical Sisterhood theorem of Gonczarowski and Friedgut. We then
recognized that the state-by-state version obtained here yields the finite
strict better-response result in Corollary~\ref{cor:finite}. Second, the model
identified the characterization of the set of outcomes supportable at
arbitrary-tanking equilibria that became Theorem~\ref{thm:support}. These
results were subsequently revised and proved within the same iterative
author--AI workflow.

The model also generated the explicit strict best-response cycle for
traditional own-list preference manipulation as a
contrast with the finite-improvement property of tanking. We
independently verified this example before retaining it. More generally, the
model assisted in searching for examples and counterexamples and in locating
related literature, while we determined which results and comparisons
were included in the manuscript.

After the sixth major draft, the model was used to generate several rounds of
simulated referee reports from different disciplinary perspectives. These
reports were development aids rather than independent external peer review.
We used them to make additional revisions and then performed further
editing independently. Consequently, substantial portions of the prose were
initially drafted or revised by the model under our repeated direction,
while the final organization, claims, emphasis, and presentation were selected
by us.

We independently verified every proof, theorem statement, example,
counterexample, and substantive claim retained in the final manuscript,
including the Tanking Sisterhood result, the supportability characterization,
and the cycling example. AI-generated arguments, computational checks,
literature suggestions, and simulated reviews were therefore used as research
and drafting aids rather than as substitutes for mathematical verification or
external peer review. We take responsibility for the final
mathematical content, citations, and presentation.


\end{document}